%% file: main.tex
\documentclass[pra,reprint,aps,superscriptaddress]{revtex4-2}
\usepackage[utf8]{inputenc}
\usepackage[T1]{fontenc}
\usepackage{epsfig}
\usepackage{mathptmx}
\usepackage{dcolumn}
\usepackage{bm}
\usepackage{siunitx}
\usepackage{graphicx}
\usepackage{times}
\usepackage{amssymb}
\usepackage{amsmath,amssymb}
\usepackage{amsfonts}
\usepackage{enumitem} 
\usepackage{afterpage}
\usepackage{babel}
\usepackage{setspace}
\usepackage[dvipsnames]{xcolor}
\usepackage{upgreek}
\usepackage{mathtools}
\usepackage{placeins}  

\usepackage{bibunits}

\usepackage{placeins}
\usepackage{capt-of}
\usepackage[pagebackref=false]{hyperref}
\usepackage[capitalise,nameinlink]{cleveref}

\usepackage[sort&compress]{natbib}

\usepackage[sort&compress]{natbib}

\setcitestyle{open={},close={},numbers,comma,super}

\usepackage[pagebackref=false]{hyperref}
\hypersetup{
    colorlinks  =   true,
    linkcolor   =   BlueViolet,
    citecolor   =   BlueViolet,
    filecolor   =   BlueViolet,
    urlcolor    =   BlueViolet}
\usepackage[capitalise,nameinlink]{cleveref}
\usepackage{hypcap}

\usepackage{titlesec}
\titleformat{\section}[hang]{\small\bfseries\sffamily}{\thesection.}{0.5em}{}
\titlespacing{\section}{0pc}{1.2pc}{0.3pc}
\titlespacing{\subsection}{0pc}{1pc}{0.2pc}

\makeatletter
\renewcommand*{\fnum@figure}{{\normalfont\bfseries \figurename~\thefigure}}
\renewcommand*{\@caption@fignum@sep}{\textbf{ | }}
\renewcommand{\thetable}{\arabic{table}}
\renewcommand*{\fnum@table}{{\normalfont\bfseries \tablename~\thetable}}
\makeatother

\begin{document}

\title{A frequency-agile microwave–optical interface for superconducting qubits}

\author{Yufeng Wu}
\altaffiliation{These authors contributed equally to this work.}
\affiliation{Department of Electrical and Computer Engineering, Yale University, New Haven, Connecticut 06511, USA}
\author{Yiyu Zhou}
\altaffiliation{These authors contributed equally to this work.}
\affiliation{Department of Electrical and Computer Engineering, Yale University, New Haven, Connecticut 06511, USA}

\author{Haoqi Zhao}
\author{Danqing Wang}
\affiliation{Department of Electrical and Computer Engineering, Yale University, New Haven, Connecticut 06511, USA}

\author{Matthew D. LaHaye}
\author{Daniel L. Campbell}
\affiliation{Air Force Research Laboratory, Information Directorate, Rome, New York 13441, USA}

\author{Hong X. Tang}
\email[Corresponding author: ]{hong.tang@yale.edu}
\affiliation{Department of Electrical and Computer Engineering, Yale University, New Haven, Connecticut 06511, USA}
\affiliation{Department of Physics and Applied Physics, Yale University, New Haven, Connecticut 06511, USA}


\begin{abstract}
Superconducting quantum processors operate at microwave frequencies in millikelvin environments, making it challenging to interconnect distant nodes using conventional microwave wiring. Coherent microwave-to-optical (M2O) transduction enables superconducting quantum networks by interfacing itinerant microwave photons with low-loss optical fiber. However, many state-of-the-art transducers provide efficient conversion only over a narrow frequency span, complicating deployment with heterogeneous superconducting devices that are detuned by gigahertz-scale offsets. Here we demonstrate a frequency-agile microwave--optical interface that overcomes this bandwidth mismatch by cascading an electro-optic M2O transducer with a multimode microwave-to-microwave (M2M) frequency converter, with \emph{in situ} tunability of the microwave resonances in both stages. Using this architecture, we realize continuous frequency coverage from 5.0 to 8.5~GHz within a single system. As an application relevant to superconducting-qubit networking, we use the cascaded M2M--M2O interface to optically read out a superconducting qubit whose readout resonator is detuned by \(1.7~\mathrm{GHz}\) from the native M2O microwave resonance, demonstrating a scalable route toward fiber-linked superconducting quantum nodes.

\end{abstract}
\maketitle
\noindent 
Superconducting quantum processors have emerged as a leading platform for quantum computation, offering fast gate operations, strong nonlinearities, and compatibility with large-scale microwave control hardware~\cite{devoret2013superconducting}. However, superconducting circuits operate at microwave frequencies in millikelvin environments, which makes it challenging to interconnect distant processors or sensors using conventional microwave wiring. Quantum networks provide a path forward by linking cryogenic superconducting nodes through low-loss optical fiber, enabling distributed quantum computing, secure quantum communication, and distributed quantum sensing~\cite{kimble2008quantum,simon2017towards,wehner2018quantum}. Realizing such networks requires a coherent microwave-to-optical (M2O) interface that maps itinerant microwave photons to optical photons and back, bridging the frequency gap between on-chip superconducting circuits and fiber-based photonic links. An efficient, low-noise, and sufficiently broadband M2O transducer is therefore a key enabling component for scaling superconducting quantum systems from local processors to geographically separated network nodes~\cite{lauk2020perspectives,han2021microwave}.

A variety of platforms are being pursued to realize coherent microwave--optical transduction, including phonon-mediated converters~\cite{han2020cavity,jiang2020efficient,mirhosseini2020superconducting,weaver2024integrated,van2025optical,yang2025multi,pan2025all}, low-frequency electromechanical approaches~\cite{higginbotham2018harnessing,arnold2020converting}, electro-optic transducers~\cite{rueda2016efficient,fan2018superconducting,holzgrafe2020cavity,mckenna2020cryogenic,fu2021cavity,xu2021bidirectional,sahu2022quantum,pintus2022ultralow,delaney2022superconducting,sahu2023entangling,zhou2025kilometer,warner2025coherent}, magnon-based hybrid systems~\cite{zhu2020waveguide,pintus2022integrated,shen2022coherent}, rare-earth-ion ensembles~\cite{fernandez2019cavity,rochman2023microwave,xie2025scalable,nicolas2023coherent}, and Rydberg-atom interfaces~\cite{han2018coherent,vogt2019efficient}. Despite rapid progress, many platforms remain constrained by limited microwave conversion bandwidth, since efficient transduction typically occurs only within a narrow frequency span set by the relevant mode linewidths. The microwave component of this conversion constraint makes a fixed-frequency transducer difficult to deploy with superconducting devices that are detuned by several gigahertz. In many high-efficiency electro-optic and optomechanical implementations, this limitation is tied to the triple-resonance requirement, where the splitting between two optical modes is engineered to match a microwave resonance. The optical doublet is often realized using coupled resonators with a splitting set by the inter-resonator coupling, and a microwave resonator is then fabricated to match this splitting~\cite{mckenna2020cryogenic}. However, the approach demands tight fabrication tolerances and typically yields a conversion window on the order of the microwave and optical linewidths, often at the \(\sim 10~\mathrm{MHz}\) scale~\cite{zhou2025kilometer}. Similarly, atomic and solid-state ensemble approaches based on Rydberg atoms or rare-earth ions are tied to specific internal transitions, which restrict the accessible microwave bands and limit frequency coverage across disparate superconducting devices~\cite{vogt2019efficient,xie2025scalable}. Traveling-wave photonic--phononic waveguides exploit broadband phase matching to extend interaction bandwidths, demonstrating an operational conversion span of \( 250~\mathrm{MHz}\)~\cite{yang2025multi}. However, this remains well below the multi-gigahertz offsets commonly encountered in heterogeneous superconducting systems, so additional frequency bridging is still needed for general GHz-scale interoperability~\cite{pan2025all}. Recent efforts to relax the fabrication requirements to match the source to the M2O transducer still only yield tens \cite{miyamura2025generation} to hundreds of MHz of conversion bandwidth \cite{campbell2026transmon}, which still require specialized superconducting circuits that are tailored to a specific M2O transducer.

Here we present a frequency-agile microwave--optical interface that addresses this bandwidth mismatch by cascading an electro-optic microwave-to-optical transducer with a multimode microwave-to-microwave (M2M) frequency converter. The M2O platform is implemented with NbN kinetic-inductance resonators and provides \emph{in situ} tunability by \(90~\mathrm{MHz}\) of the microwave resonances~\cite{zhou2025kilometer}. To extend far beyond this range, we incorporate an NbN kinetic-inductance multimode ring resonator that performs M2M conversion~\cite{wu2025broad}. The ring supports a dense set of resonant modes with a free spectral range (FSR) of 76~MHz, and its modal frequencies are likewise flux-tunable. The combined tunability of the M2O operating point and the M2M mode spectrum provides two independent degrees of freedom for frequency alignment. When the M2M signal mode is tuned into resonance with a target microwave device, the M2O microwave resonance can be tuned to match the corresponding M2M idler, enabling M2M-mediated M2O transduction. Using this architecture, we demonstrate frequency coverage from 5.0 to 8.5~GHz within a single reconfigurable system.

To illustrate the utility of this frequency-agile interface in a superconducting-circuit setting, we employ the cascaded M2M--M2O system to optically readout a superconducting qubit whose readout resonator is detuned by \(1.7~\mathrm{GHz}\) from the M2O microwave resonance. Optical access to superconducting circuits is attractive for scalable cryogenic systems because fiber links provide low thermal conductivity, strong isolation, and a compact routing footprint relative to extensive microwave wiring, while reducing the heat load associated with multiple microwave lines and cryogenic amplification stages~\cite{delaney2022superconducting,arnold2025all,pan2025all,van2025optical}. In our architecture, the drive frequency is determined by the qubit readout resonator, and the M2M stage bridges the frequency gap by translating this signal into the M2O conversion band, enabling optical detection without tuning the superconducting device.

\section*{Device, experimental schematic and frequency matching}
\label{sec:freq_matching}

\begin{figure}[t]
\capstart
\centering
\includegraphics[width=\linewidth]{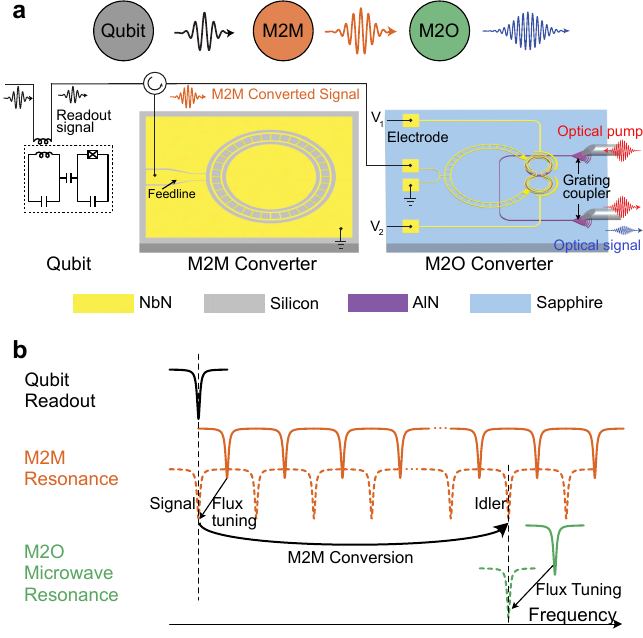}
\caption{\label{fig:concept} %
\textbf{Concept of a cascaded M2M--M2O microwave--optical interface with deterministic frequency matching.}
\textbf{(a)} Frequency-agile qubit readout via cascaded conversion. A fixed-frequency qubit readout tone is first sent to a multimode M2M converter, where an applied M2M pump enables coherent translation from a selected signal mode to a nearby idler mode. The translated microwave tone is then routed to an M2O transducer and converted to an optical sideband under optical pumping.
\textbf{(b)} Deterministic frequency-matching protocol. The qubit readout frequency (black, fixed) is initially detuned from the M2O microwave resonance (green). The M2M spectrum (orange) is flux-tuned to bring a signal mode into resonance with the readout tone. The M2O microwave resonance is then flux-tuned to the nearest M2M mode, designated as the idler. With the M2M pump applied, the readout signal is coherently converted from the signal mode to the idler mode, bridging the frequency gap and enabling subsequent microwave-to-optical transduction in the M2O stage.}
\end{figure}

As shown in Fig.~\ref{fig:concept}(a), the M2M device is patterned from a \(5~\mathrm{nm}\)-thick NbN film on a \(500~\mu\mathrm{m}\)-thick high-resistivity silicon substrate. For clarity it is depicted as a ring resonator, while the realized structure is a compact meander line with a total physical length of \(80~\mathrm{mm}\). Internal loop sections with asymmetric linewidth provide flux tunability and enable three-wave mixing \cite{wu2025broad}. The M2M converter is configured as a single-port reflective device: the input signal and pump are applied at a single port, and the generated idler, together with the reflected tones, returns through the same port.

The M2O device co-integrates optical and microwave components on a sapphire substrate. The optical subsystem is implemented in AlN as an evanescently coupled double-ring resonator, with light coupled on and off chip through a pair of grating couplers \cite{zhou2025kilometer}. The microwave subsystem is a \(50~\mathrm{nm}\)-thick NbN resonator that incorporates the same loop geometry as the M2M device, enabling \emph{in situ} frequency tunability.

Figure~\ref{fig:concept}(a) illustrates the readout architecture. A fixed-frequency microwave tone, encoded with qubit readout information, is first routed to the multimode M2M stage. Here, it is frequency-translated to match the operating range of the M2O device. This translated tone is then delivered to the M2O device and converted into an optical sideband via a strong optical pump. This architecture enables high-fidelity optical detection of the qubit state without requiring any retuning of the superconducting circuit.

\begin{figure*}[t]
\capstart
\centering
\includegraphics[width=\linewidth]{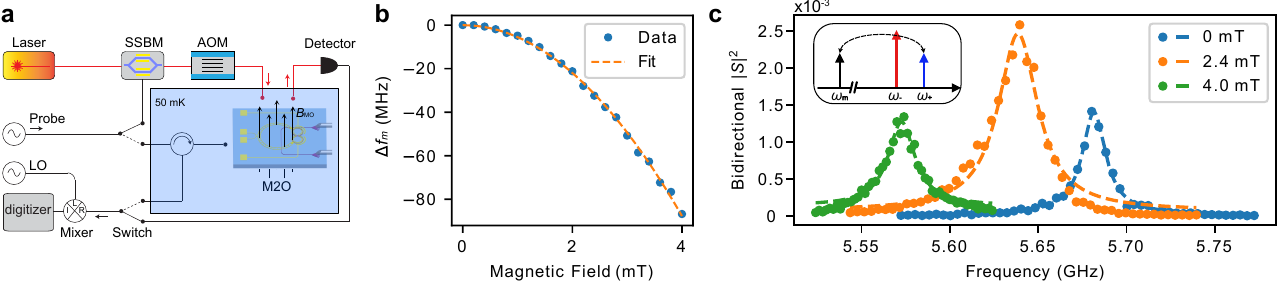}
\caption{\textbf{Transducer characterization.} \textbf{(a)} Simplified schematic of the setup to characterize the transduction efficiency. SSBM, single sideband modulator; AOM, acousto-optic modulator; LO, local oscillator. 
\textbf{(b)} Magnetic-field tuning of the microwave resonance. Measured frequency shift $\Delta f_{\mathrm{m}}$ (relative to the zero-field resonance) as a function of applied magnetic field. \textbf{(c)} Calibrated bidirectional electro--optic conversion efficiency spectrum $|S|^{2}$ at three magnetic-field bias points (0, 2.4, and 4.0~mT). Inset: pumping scheme with a strong optical pump at $\omega_-$ enables microwave to blue-sideband light conversion.}
\label{Fig:Seo}
\end{figure*}

Figure~\ref{fig:concept}(b) summarizes the deterministic frequency-matching protocol. The M2M spectrum is flux-tuned to bring a selected signal mode into resonance with the fixed qubit readout frequency. The M2O microwave resonance is then flux-tuned to the nearest M2M mode, which serves as the idler. Applying an M2M pump enables coherent frequency conversion between the signal and idler modes, transferring the readout signal to the idler frequency matched to the M2O resonance. Once aligned to the M2O conversion band, the translated microwave tone is transduced to an optical sideband under optical pumping in the M2O stage.
 
Compared with Josephson parametric converter (JPC) mediated approaches~\cite{pan2025all}, the cascaded M2M--M2O architecture offers two key advantages. First, the M2M device is intrinsically multimode, providing multiple signal--idler pairs and therefore substantially broader frequency coverage. This enables deterministic bridging between a wide range of fixed input frequencies and a downstream M2O resonance. In contrast, a JPC typically supports only a single signal--idler mode pair, which restricts the accessible frequency span when interfacing arbitrary inputs to a fixed target resonance. Second, the frequency-bridging M2M module can be mounted on a higher-temperature stage because NbN has a relatively high critical temperature (\(\sim 10~\mathrm{K}\)) \cite{cheng2019superconducting}, allowing operation on an elevated plate of the dilution refrigerator. Under radiative-cooling conditions, the conversion performance can remain largely unchanged~\cite{xu2024radiatively}, while reducing pump-power dissipation and the associated heat load delivered to the mixing chamber (MXC). By comparison, JPC-based conversion is typically operated at the mixing chamber and requires pump delivery and dissipation at the coldest stage, continuing to impose heat load and wiring/packaging overhead. This makes it less compatible with the central motivation of optical readout---namely reducing the footprint and thermal burden at the mixing chamber.


\section*{Characterization of tunable M2O transducer}
To calibrate the M2O transduction efficiency, we use the setup in Fig.~\ref{Fig:Seo}(a) and perform a full two-port electrical-optical characterization. The M2O transducer is mounted on the MXC stage, which is heated to 50~mK to provide increased cooling power. A microwave tone is switched either to the device microwave port through the dilution-fridge input line (electrical drive) or to an optical single-sideband modulator (SSBM) to generate an optical single sideband on a strong optical local oscillator (LO), which is injected into the device optical port (optical drive). The output is switched to either the fridge microwave output chain (electrical readout) or to heterodyne detection with the same optical LO on a fast photodetector (optical readout). Combining the two drive paths and two readout paths yields the complex scattering matrix $\{S_{\mathrm{oo}}, S_{\mathrm{oe}}, S_{\mathrm{eo}}, S_{\mathrm{ee}}\}$, from which we extract the calibrated bidirectional conversion efficiency~\cite{andrews2014bidirectional, fan2018superconducting, fu2021cavity, xu2021bidirectional, zhou2025kilometer}
\begin{equation}
\eta_\mathrm{MO} = \frac{S_{\mathrm{oe,pk}}\, S_{\mathrm{eo,pk}}}{S_{\mathrm{oo,bg}}\, S_{\mathrm{ee,bg}}},
\label{eq:eta}
\end{equation}
where ``pk'' and ``bg'' denote the on-resonance peak and off-resonance background, respectively.

To tune the microwave resonance of the M2O transducer, the device is placed inside a superconducting coil. An external magnetic field $B_{\mathrm{MO}}$ is generated by driving a current through the coil, as illustrated in the inset of Fig.~\ref{Fig:Seo}(b). Owing to the loop-like geometry of the NbN resonator, the applied flux induces a dc screening supercurrent $I_{\mathrm{sc}}$. This screening current increases the kinetic inductance according to
\begin{equation}
L_{\mathrm{k}}(I_{\mathrm{sc}})=L_{\mathrm{k0}}\!\left[1+\left(\frac{I_{\mathrm{sc}}}{I^{*}}\right)^{2}\right],
\end{equation}
where $L_{\mathrm{k0}}$ is the zero-current kinetic inductance and $I^{*}$ is the characteristic current that is on the order of the critical current. The resulting increase in total inductance shifts the microwave resonance frequency $f_{\mathrm{m}}$, which is well approximated by a quadratic dependence~\cite{xu2019frequency},
\begin{equation}
\label{equ:frequency_shift}
\frac{\Delta f_{\mathrm{m}}}{f_{\mathrm{m}}} \approx -k\,B_{\mathrm{MO}}^{2},
\end{equation}
where $k$ is set by the inductance participation ratio, $I^{*}$, and the loop geometry. Figure~\ref{Fig:Seo}(b) shows the magnetic-field dependence of the microwave resonance frequency. We plot the frequency shift $\Delta f_{\mathrm{m}} \equiv f_{\mathrm{m}}(B)-f_{\mathrm{m}}(0)$ as a function of the applied field $B_{\mathrm{MO}}$, where the zero-field resonance is $f_{\mathrm{m}}(0)=5.669~\mathrm{GHz}$. The resonance red-shifts monotonically with increasing field and reaches a total shift of $90~\mathrm{MHz}$ (1.6\% of $f_{\mathrm{m}}(0)$) at $B_{\mathrm{MO}}=4~\mathrm{mT}$. The dashed curve is a fit to the expected quadratic dependence in Eq. \ref{equ:frequency_shift}. This tuning range already exceeds the free spectral range (FSR) of the M2M device (76~MHz), enabling the M2O resonance to be positioned anywhere within one full M2M mode spacing and thus cover an entire FSR-wide band of the M2M spectrum. 

\begin{figure*}[t]
\capstart
\centering
\includegraphics[width=\linewidth]{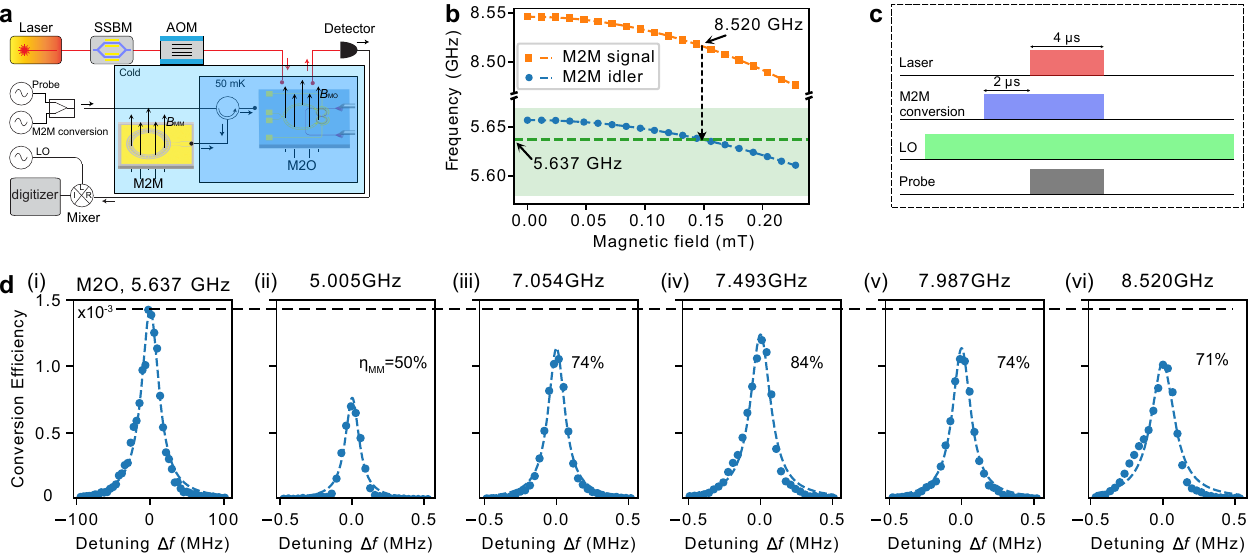}
\caption{\textbf{M2M--M2O cascaded conversion setup and performance.}
\textbf{(a)} Experimental schematic of the cascaded M2M--M2O measurement. A microwave signal is first frequency-translated by the M2M stage and then converted to the optical domain by the M2O transducer.
\textbf{(b)} Flux tuning of the M2M signal and idler frequencies relative to the fixed M2O operating frequency at 5.637~GHz. The shaded region indicates the accessible M2O tuning range.
\textbf{(c)} Pulse diagram of the measurement sequence.
\textbf{(d)} Calibrated conversion-efficiency spectra. Panel (i) shows the baseline M2O-only conversion efficiency at 5.637~GHz, which sets the reference performance (indicated by the dashed line). Panels (ii--vi) show the cascaded M2M--M2O conversion efficiency at several translated microwave signal frequencies from 5.0 to 8.5 GHz.
}
\label{Fig3:M2M_M2O}
\end{figure*}

Efficient electro--optical conversion also requires three-mode frequency matching between the microwave resonance and the optical mode pair:
$\omega_{+}-\omega_{-}=\omega_{m}$,
where $\omega_{+}$ and $\omega_{-}$ denote the blue- and red-detuned optical modes, respectively. At zero bias, the optical mode splitting is $\omega_{+}-\omega_{-}=2\pi\times 5.354~\mathrm{GHz}$, which is detuned from the microwave frequency by $\sim 300~\mathrm{MHz}$. We compensate this mismatch using dc electrostatic tuning, with a differential voltage applied across the two on-chip electrodes ($+160$~V and $-160$~V), bringing the optical splitting into resonance with $\omega_m$ \cite{zhou2025kilometer}. 

Figure~\ref{Fig:Seo}(c) shows the calibrated bidirectional conversion efficiency spectrum \(|S|^{2}\), measured at three magnetic-field bias points (0, 2.4, and 4.0~mT). Shown in the inset is the pumping scheme: a strong optical pump at frequency \(\omega_{-}\) enables coherent conversion between the microwave field and the blue-sideband optical mode. In the measurement, the optical pump with a 10 dBm on-chip power is applied in 4~$\mu$s pulses repeated every 1~ms. As the magnetic field is increased, the conversion peak shifts to lower frequency, closely following the flux-tuned microwave resonance of the NbN loop resonator. The peak efficiency remains on the order of \(10^{-3}\) throughout the tuning range, indicating that flux tuning repositions the transduction window without substantially degrading the conversion strength. The measured 3-dB conversion bandwidths are 16.9~MHz, 31.4~MHz, and 22.6~MHz, with the variation primarily arising from differences in the microwave--optical frequency-matching condition at each bias point.

\section*{Broadband tuning via cascaded M2M--M2O conversion}
The conversion-frequency tunability in the standalone M2O device (Fig.~\ref{Fig:Seo}(c)) is limited by the accessible flux-induced shift of the microwave resonance (\(<100~\mathrm{MHz}\)). To extend the usable spectral coverage, we cascade an M2M frequency converter in front of the M2O transducer, implementing an M2M--M2O conversion chain as illustrated in Fig.~\ref{Fig3:M2M_M2O}(a). In this architecture, the incoming microwave signal is first translated by the M2M stage to the fixed operating frequency of the M2O device, and is subsequently converted to the optical domain by the electro--optic transducer. 

\begin{figure*}[t]
\capstart
\centering
\includegraphics[width=\linewidth]{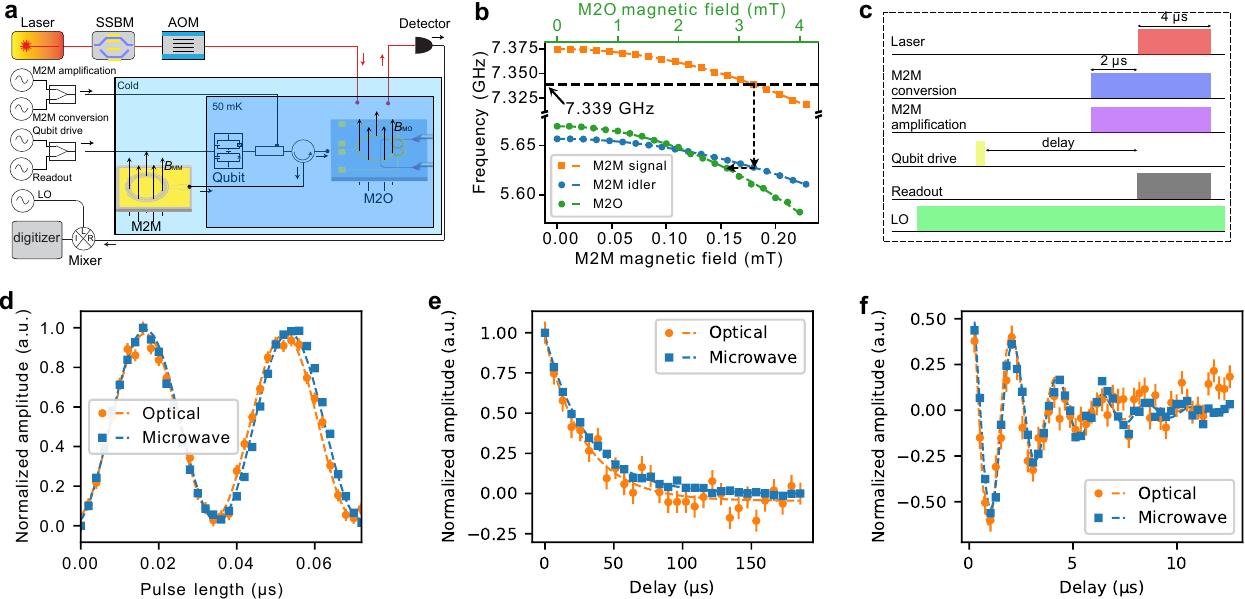}
\caption{\textbf{Optical readout of a superconducting qubit using a cascaded M2M--M2O link.}
\textbf{(a)} Measurement schematic: the qubit readout tone is routed through an M2M frequency converter (providing frequency translation and gain) before entering the M2O transducer for optical detection.
\textbf{(b)} Deterministic frequency matching among the qubit readout, the M2M mode pair, and the M2O microwave resonance. Flux tuning aligns the M2M signal mode (orange) with the fixed qubit readout frequency (orange dashed line) while shifting the corresponding idler mode (blue). The M2O microwave resonance (green) is then tuned to coincide with the idler.
\textbf{(c)} Pulse sequence for the qubit $T_1$ experiment with optical readout.
\textbf{(d-f)} Rabi oscillations, $T_1$, and $T_2$ measurements with optical (orange) and conventional microwave (blue) readout.}
\label{Fig4:readout}
\end{figure*}

The M2M module is mounted on the cold plate to confine pump-induced dissipation to higher-temperature stages and minimize heating of the MXC plate. The M2O device operates at a fixed microwave frequency of 5.637~GHz. We tune the M2M resonance using a second coil (with magnetic field $B_{\mathrm{MM}}$) that is magnetically isolated from the M2O device, enabling independent control without measurable crosstalk, and align the M2M idler frequency to the M2O resonance, as shown in Fig.~\ref{Fig3:M2M_M2O}(b). The input signal (5.0--8.5~GHz) is combined with an M2M pump tone whose frequency is set to the signal--idler difference, enabling frequency translation into the M2O band. Fig.~\ref{Fig3:M2M_M2O}(c) shows the pulse sequence: the M2M pump is applied \(2~\mu\text{s}\) before the laser and probe to initiate and stabilize the M2M conversion. The measurement window is \(4~\mu\text{s}\), repeated with a 1~ms period.

With the cascaded configuration, the effective transduction window can be repositioned over a multi-gigahertz span without requiring large \emph{in situ} tuning of the M2O resonance. Figure~\ref{Fig3:M2M_M2O}(d) presents representative conversion spectra at several translated microwave frequencies. A full two-port electro--optical scattering-matrix calibration of the cascaded chain is not available because an \emph{in situ} O2M--M2M configuration cannot be implemented. Instead, we define the overall efficiency as the product of the independently measured stage efficiencies,
\begin{equation}
\eta = \eta_{\mathrm{MO}}\,\eta_{\mathrm{MM}},
\end{equation}
where \(\eta_{\mathrm{MO}}\) and \(\eta_{\mathrm{MM}}\) are obtained from separate two-port calibrations of the M2O and M2M devices \cite{wu2025broad}. As shown in Fig.~\ref{Fig3:M2M_M2O}(d), the M2O stage maintains an efficiency on the order of \(10^{-3}\), while the M2M stage provides high translation efficiency, with \(\eta_{\mathrm{MM}} > 50\%\) for all operating modes. In this demonstration, we show five representative modes spanning 5.0--8.5~GHz for the cascaded conversion. As demonstrated in Ref.~\cite{wu2025broad}, this approach can be extended to convert any accessible resonant mode within this frequency range.

\section*{Optical readout of a detuned superconducting qubit}
To demonstrate the utility of the cascaded converter, we optically readout a superconducting qubit whose readout resonator is at 7.339~GHz, \(\sim 1.7~\)GHz detuned from the microwave resonance (5.669~GHz) of the standalone M2O device. Fig.~\ref{Fig4:readout}(a) shows the measurement schematic, where the qubit--resonator output is routed into the M2M stage prior to the M2O transducer. In this configuration, the M2M converter bridges the frequency gap by translating the qubit readout tone to an idler frequency that lies within the M2O conversion band, enabling optical qubit readout.

Figure~\ref{Fig4:readout}(b) summarizes the frequency-matching procedure. First, the M2M signal mode is tuned into resonance with the qubit readout tone at 7.339~GHz to maximize coupling to the outgoing qubit field. As the M2M device is flux tuned, the corresponding idler mode is simultaneously shifted into the M2O operating band. The M2O device is then flux biased to align its microwave resonance with the M2M idler frequency, thereby satisfying the resonance condition for the cascaded M2M--M2O link and maximizing the overall optical readout efficiency.

In the readout experiment, the M2O device is driven with 0~dBm on-chip optical pump power, yielding a conversion efficiency of $1.3\times 10^{-4}$. This reduced pump power is intended to minimize backaction on both the qubit and the M2M stage; notably, recent work suggests that immersing superconducting circuits in superfluid helium can mitigate such effects \cite{li2025superfluid}. This relatively low M2O conversion efficiency motivated us to demonstrate the versatility of the cascaded architecture by operating the M2M device in a combined conversion-and-amplification mode. Specifically, an M2M amplification tone at twice the idler frequency is applied simultaneously with the M2M conversion pump, enabling frequency translation while providing gain. Under these conditions, the M2M stage operates as a phase-sensitive amplifier, providing $30$ dB of gain with a signal-to-idler internal conversion efficiency of  $67\%$. 

The pulse sequence for optical qubit readout is shown in Fig.~\ref{Fig4:readout}(c) for a representative $T_{1}$ measurement. After preparing the qubit in the excited state with a qubit-drive pulse, we wait for a variable delay. As in our previous protocol, we apply the M2M conversion- and amplification-pump pulses $2~\mu\text{s}$ before the readout to turn on and stabilize the conversion-and-gain process. We then apply a $4~\mu\text{s}$ microwave readout pulse to the qubit together with a simultaneous $4~\mu\text{s}$ optical pump for the M2O transducer; the resulting readout field is frequency-converted and amplified by the M2M stage and simultaneously upconverted to the optical domain.


In Fig.~\ref{Fig4:readout}(d--f), we benchmark the cascaded optical readout against the conventional microwave readout using standard qubit characterization measurements. The Rabi oscillations obtained by sweeping the drive pulse length [Fig.~\ref{Fig4:readout}(d)] show close agreement in both cases, showing that the optical readout reproduces the expected qubit dynamics under coherent driving. We then extract the energy-relaxation time by preparing the qubit in \(|e\rangle\), waiting for a variable delay, and measuring the excited-state population [Fig.~\ref{Fig4:readout}(e)]. Exponential fits yield \(T_{1}^{\mu\mathrm{w}} = 30.9 \pm 0.3~\mu\text{s}\) for microwave readout and \(T_{1}^{\mathrm{opt}} = 28.7 \pm 3.1~\mu\text{s}\) for optical readout, consistent within the quoted uncertainties. Finally, Ramsey measurements [Fig.~\ref{Fig4:readout}(f)] give \(T_{2}^{\mu\mathrm{w}} = 3.1 \pm 0.1~\mu\text{s}\) and \(T_{2}^{\mathrm{opt}} = 2.6 \pm 0.3~\mu\text{s}\). The relatively low $T_2$ was limited by the thermal noise in the readout resonator caused by the elevated operating temperature of the MXC \cite{clerk2007using, rigetti2012superconducting}. Overall, the optical readout reproduces the key coherence metrics with only a modest degradation and increased error bars, validating the cascaded M2M--M2O link for quantitative qubit measurements.

Throughout the measurements, both the M2O and M2M devices remain near their ground states. The M2O microwave-mode temperature is calibrated using a variable-temperature stage (VTS) that provides a known thermal-noise reference (see Supplementary Information for details), yielding an occupancy of $3.1 \pm 0.2$ quanta, consistent with prior measurements on similar platforms~\cite{fu2021cavity, zhou2025kilometer}. Under operating conditions, the M2M mode temperature is independently calibrated to be below $1.3$ quanta (see Supplementary Information). We attribute the residual degradation in optical-readout performance primarily to the limited M2O conversion efficiency. 

\section*{Conclusions and outlook}
To conclude, we demonstrate a superconducting microwave--optical interface that combines a flux-tunable microwave-to-optical transducer with a multimode microwave-to-microwave frequency-conversion stage, extending the usable conversion bandwidth and enabling optical readout of detuned superconducting qubits. A central requirement is deterministic frequency alignment, i.e., the M2M idler can be tuned to the target frequency to be converted to optical light. In our current implementation, the M2M tuning range is limited to \(0.83\%\), corresponding to a \(78\%\) frequency coverage around the 7.339~GHz target frequency~\cite{wu2025broad}. This limitation can be mitigated by reducing the M2M FSR, for example by fabricating a longer ring resonator.

It is possible for the cascaded M2M--M2O link to support multiplexed qubit readout, provided the readout-resonator frequencies can be tuned over approximately one M2M free spectral range (FSR). Such tuning has been demonstrated previously~\cite{wang2013quantum}.
In a frequency-multiplexed architecture, each qubit couples to a distinct readout resonator, and the converter must provide frequency matching for multiple channels. In our approach, multiplexing can be achieved by assigning each resonator to a different M2M signal mode and tuning the resonator to the nearest available M2M mode within one FSR. This requirement is substantially less stringent than the multi-gigahertz tuning that would be needed to directly interface multiple fixed-frequency readout resonators to an M2O device with a fixed microwave resonance.

Beyond readout, optical control of superconducting qubits is an important next step toward reducing microwave-induced heat load and improving scalability to large qubit numbers. Several recent works have demonstrated optical-to-microwave (O2M) links and optically mediated control primitives in superconducting platforms~\cite{lecocq2021control,warner2025coherent,van2025optical,pan2025all}. In our framework, the same hardware can be operated in reverse as an O2M--M2M chain for qubit control: the O2M stage converts modulated optical fields into microwave tones near its resonance, and the M2M stage then bridges the frequency gap between the M2O resonance and the target qubit frequency. This reverse-operation architecture also naturally supports multiplexing. A single O2M-generated microwave tone can be aligned to an M2M signal mode, and by applying multiple M2M conversion pumps, it can be translated into multiple idler frequencies, each matched to a different qubit. This bidirectional capability provides a unified route to both optical readout and optical control, enabling frequency-agile optical interconnects for scalable superconducting quantum processors.

\def\bibsection{\section*{References}}
\bibliographystyle{modified.bst}

\bibliography{My_reference}
\clearpage

\section*{Methods}
\noindent \textbf{Device fabrication and packaging}\\
Fabrication of the M2O device begins with a commercially available 1~\textmu m-thick aluminum nitride (AlN) template on a sapphire substrate. A silicon dioxide (SiO$_2$) hard mask is first deposited by plasma-enhanced chemical vapor deposition (PECVD). For the first patterning step, an electron-beam resist (CSAR 62) is spin-coated, exposed using an EBPG 5200 e-beam lithography system, and developed in xylene. The SiO$_2$ hard mask is etched in a CHF$_3$/O$_2$ plasma, followed by an inductively coupled plasma etch of 400~nm of AlN using a Cl$_2$/BCl$_3$/Ar chemistry. The remaining SiO$_2$ mask is removed in buffered oxide etchant (BOE). A second SiO$_2$ hard mask is then deposited by PECVD for the next lithography round. In this step, a negative-tone e-beam resist (Ma-N 2405) is used to define the pattern, and the SiO$_2$ hard mask is again etched in CHF$_3$/O$_2$. The remaining 600~nm of AlN is subsequently etched using Cl$_2$/BCl$_3$/Ar, completing the full 1~\textmu m AlN etch. Afterward, a 2.5~\textmu m-thick SiO$_2$ cladding layer is deposited by PECVD, and the chip is annealed at 1000~$^\circ$C for 2~hours. A 50~nm-thick niobium nitride (NbN) film is then deposited as the superconducting layer by atomic layer deposition (ALD). The NbN is patterned using photolithography with Shipley S1805 resist, exposed using a Heidelberg MLA150 maskless aligner, and developed in MF-319. The NbN is etched in a CF$_4$ plasma to define the microwave circuits. For optical access, a fiber array is aligned and glued to the chip to interface with the grating couplers, with an insertion loss of 5 dB per facet. The on-chip electrodes are wire-bonded to a printed circuit board (PCB), and the assembly is enclosed in a copper package to suppress stray optical and microwave radiation. The copper enclosure also provides electromagnetic isolation, reducing the impact of the optical illumination and dc bias fields used for M2O operation on nearby superconducting qubit circuitry.

The M2M device is fabricated on a 500~\textmu m-thick high-resistivity silicon substrate. A 5~nm niobium nitride (NbN) film is deposited by ALD, yielding a sheet inductance of \(105~\mathrm{pH}/\square\). The microwave circuit is defined in a single electron-beam lithography step. Briefly, a CSAR resist is spin-coated to serve as an etch mask, patterned by e-beam exposure, and developed. The exposed NbN is then etched in a CF$_4$ plasma, transferring the lithographic pattern into the superconducting film. The M2M chip is mounted in an oxygen-free copper package and wire-bonded to a PCB for access to the microwave signal.

\noindent \textbf{Theory on M2O conversion}\\
The derivation of the electro-optical single-photon coupling rate $g_{\text{eo}}$ follows our previous work \cite{zhou2025kilometer}. For a double-ring resonator, the system Hamiltonian can be written as 
\begin{align}
\begin{split}
H = & \hslash \omega_- a_{-}^{\dagger} a_{-}  + \hslash \omega_+ a_{+}^{\dagger} a_{+}^{} + \hslash \omega_{\text{m}} a_{\text{m}}^{\dagger} a_{\text{m}} \\
&+ \hslash g_{\text{eo}}(   a_{\text{m}}  a_{-}  a_{+}^{\dagger}   + a_{\text{m}}^{\dagger}  a_{-}^{\dagger}  a_{+} ),
\end{split}
\end{align}
where $a_{-}$ ($\omega_{-}$) is the annihilation operator (eigenfrequency) of the optical red-sideband mode, $a_{+}$ ($\omega_{+}$) is the annihilation operator (eigenfrequency) of the optical blue-sideband mode, and $a_{\text{m}}$ ($\omega_{\text{m}}$) is the annihilation operator (eigenfrequency) of the microwave mode. $g_{\text{eo}}$ is the electro-optical single-photon coupling rate, which can be explicitly expressed as
\begin{align}
\begin{split}
g_{\text{eo}} = & \sqrt{  \frac{ \hslash \omega_{\text{-}}  \omega_{\text{+}}   \omega_{\text{m}}   }{8 \pi \epsilon_0 R}} \frac{\iint drdz \: \epsilon_{\text{o,zz}}^2 r_{33}  |u_{\text{o,z}}|^2 u_{\text{m,z}}    }{\iint drdz \epsilon_{\text{o,zz}} |u_{\text{o,z}}|^2   } \\
& \cdot \frac{1}{\sqrt{\iint  (\epsilon_{\text{m,rr}} |u_{\text{m,r}}|^2 + \epsilon_{\text{m,zz}} |u_{\text{m,z}}|^2) drdz }},
\end{split}
\end{align}
where $R$ is the AlN ring radius, $\epsilon_0$ is the vacuum permittivity, $\hslash$ is the reduced Planck constant, $u_{\text{m,r}}$ and $u_{\text{m,z}}$ are the $r$ and $z$ component of the transverse spatial profile of the microwave mode, and $u_{\text{o,z}}$ is the $z$ component of the transverse spatial profile of optical modes. Because the spatial profiles of both optical red-sideband and blue-sideband modes are nearly identical, we use $u_{\text{o,z}}$ to denote the profile of both modes. $\epsilon_{\text{o,zz}}$ is the relative permittivity for optical fields along the $z$ direction, $\epsilon_{\text{m,zz}}$ ($\epsilon_{\text{m,rr}}$) is the relative permittivity for microwave fields along the $z$ ($r$) direction, and $r_{33} = 1$~pm/V is the electro-optic coefficient of AlN. The transverse spatial profiles of optical and microwave modes can be calculated by a commercial mode solver, and the numerical result is $g_{\text{eo}}=2\pi\cdot 264$~Hz, which is close to the experimental value $g_{\text{eo}}\approx 2\pi\cdot 272$~Hz.

When an optical pump of power $P_o$ is tuned to the red-sideband resonance frequency, the intra-cavity pump photon number is $n_{p}=  4 P_o \kappa_{-,\text{ex}} /( \hslash \omega_{-} \kappa_{-}^{2})$. The on-chip transduction efficiency can be calculated as 
\begin{equation}
\eta_{\text{eo}}= \frac{\kappa_{+,\text{ex}} }{ \kappa_{+} }  \frac{ \kappa_{\text{m,ex}} }{ \kappa_{\text{m}}  } \frac{4C_{\text{eo}}}{(1+C_{\text{eo}})^2},
\label{eq:efficiency}
\end{equation}
where $C_{\text{eo}}=4 n_p g_{\text{eo}}^2/(\kappa_{\text{m}} \kappa_{+})$ is the electro-optic cooperativity, $\kappa_{i,\text{ex}}$ ($\kappa_{i,\text{in}}$) denotes the external (intrinsic) coupling rate of mode $i$, where $i=\{-,+,
\text{m}\}$ corresponds to the optical red-sideband mode, blue-sideband mode, and microwave mode, respectively. $\kappa_i = \kappa_{i,\text{in}} + \kappa_{i,\text{ex}}$ denotes the total coupling rate for each mode.

\noindent \textbf{Theory on M2M conversion and amplification}
\label{sec:methods_effH_rot}\\
We describe the M2M stage using two microwave modes, a passive signal mode $a$ with resonance frequency $\omega_a$, and an idler mode $b$ with resonance frequency $\omega_b$. Mode conversion between $a$ and $b$ is enabled by a pump-assisted nonlinear interaction that produces an effective beam-splitter coupling. In addition, the idler mode is parametrically driven by an amplification pump at approximately twice its resonance frequency, $\omega_{p,\mathrm{amp}}\simeq 2\omega_b$, so that $b$ acts as a degenerate parametric amplifier (DPA). Treating the pumps as classical tones and working in the linearized small-signal regime, we move to a rotating frame at frequencies $\omega_{ra}$ and $\omega_{rb}$ for modes $a$ and $b$, respectively. In this frame, the effective Hamiltonian takes the time-independent form
\begin{equation}
\label{eq:Heff_rot}
\begin{split}
\frac{H_{\mathrm{eff}}}{\hslash}
= & 
-\Delta_a\, a^\dagger a
-\Delta_b\, b^\dagger b
+
\left(g\,e^{-i\phi_g}\, a b^\dagger + g\,e^{+i\phi_g}\, a^\dagger b\right) \\
 & +
\frac{\epsilon}{2}\left(e^{-i\phi_\epsilon}\, b^2 + e^{+i\phi_\epsilon}\, b^{\dagger 2}\right),
\end{split}
\end{equation}
where $\Delta_a=\omega_{ra}-\omega_a$ and $\Delta_b=\omega_{rb}-\omega_b$ are the detunings in the rotating frame, $g$ is the (pump-dependent) conversion rate with phase $\phi_g$, and $\epsilon$ is the parametric pump strength on mode $b$ (set by the $2\omega_b$ pump) with phase $\phi_\epsilon$. 

In the absence of the parametric term ($\epsilon=0$), Eq.~\eqref{eq:Heff_rot} reduces to the standard two-mode frequency-conversion Hamiltonian. In this limit the interaction is purely beam-splitter-like and conserves excitation number, implementing coherent state transfer between the two modes. The conversion dynamics are governed by the dimensionless conversion cooperativity
\begin{equation}
C \equiv \frac{4g^2}{\kappa_a \kappa_b},
\end{equation}
where $\kappa_{a(b)}$ are the total energy decay rates of modes $a(b)$. For an ideal lossless converter operated on resonance, the internal conversion efficiency is maximized near the impedance-matching condition $C\simeq 1$, corresponding to $|g|\sim \sqrt{\kappa_a\kappa_b}/2$.

When the idler mode is additionally parametrically driven ($\epsilon\neq 0$), mode $b$ operates as a degenerate parametric amplifier: it provides parametric gain and generates squeezed-vacuum fluctuations. In the linearized, undepleted-pump regime considered here, the resulting M2M stage remains a linear network, in the sense that the output fields are linear (Bogoliubov) transformations of the input fields. Although the parametric drive acts only on $b$, the beam-splitter coupling $g$ coherently exchanges excitations between the two modes, so the pump-modified response of $b$ feeds back onto the signal mode. This backaction does not introduce nonlinearity; rather, it renormalizes the effective susceptibility seen by mode $a$ and thereby modifies the conversion transfer function. Equivalently, one may view mode $a$ as being coupled to an active idler bath whose pump-dependent susceptibility and amplified fluctuations are partially mapped onto the signal mode through $g$.


\section*{{\normalsize{}Acknowledgments}}
\noindent\footnotesize This work is funded by Co-design Center for Quantum Advantage (DE-SC0012704). HXT acknowledges funding by the Office of Naval Research (ONR) under award number N00014-23-1-2121 and Air Force Office of Scientific Research (AFOSR) under award number FA9550-23-1-0338. Y.Z. acknowledges the support from Yale Quantum Institute fellowship. The authors would like to thank Yong Sun, Lauren McCabe, Kelly Woods, Yeongjae Shin, Michael Rooks, and Sungwoo Sohn for their assistance provided in the device fabrication. The fabrication of the devices was done at the Yale School of Engineering \&
Applied Science (SEAS) Cleanroom and the Yale Institute for Nanoscience and Quantum Engineering (YINQE). The authors would like to thank Yuvraj Mohan and the Rigetti Quantum Foundry Services team for design support and device fabrication of the transmon chip studied in this work.

\section*{{\normalsize{}Author contributions}}
\noindent\footnotesize H.X.T. and Y.W. conceived the idea and experiment; Y.W. fabricated the M2M device; Y.Z. fabricated the M2O device; D.L.C. and M.D.L. designed the qubit device; Y.W. and H.Z. performed the measurements with assistance from Y.Z., D.W. and D.L.C.; Y.W., H.Z., and
H.X.T. analyzed the data; Y.W. and H.X.T. wrote the manuscript with inputs
from all authors; H.X.T. supervised the project.

\section*{{\normalsize{}Competing interests}}
\noindent \footnotesize The authors declare no competing interests.

\section*{{\normalsize{}Additional information}}
{\noindent\footnotesize{}\textbf{Correspondence and requests for materials} should be addressed to H.X.T.}
{\footnotesize\par}

\clearpage
\normalsize
\input{supplement}


\end{document}

%% file: supplement.tex

\clearpage
\FloatBarrier
\onecolumngrid

\renewcommand{\thefigure}{S\arabic{figure}}
\setcounter{figure}{0}
\renewcommand{\thetable}{S\arabic{table}}
\setcounter{table}{0}

\begin{center}
{\Large\bfseries Supplementary Material for}\\[2pt]
{\Large\bfseries A frequency-agile microwave--optical interface for superconducting qubits}
\end{center}

\section{Summary of experimentally measured system parameters}

\begin{center}
\renewcommand{\arraystretch}{1.2}

\captionof{table}{\textbf{Experimentally measured system parameters.}}
\label{table:parameter}

\begin{tabular}{|l|c|}
\hline
\multicolumn{1}{|c|}{M2O Parameters} & Value \\ \hline
Optical red sideband mode frequency $\omega_{+}$ &
$2\pi \cdot 194.1069~\mathrm{THz}$ \\ \hline
Optical red sideband mode intrinsic loss rate $\kappa_{+,\mathrm{in}}$ &
$2\pi \cdot 165.7~\mathrm{MHz}$ \\ \hline
Optical red sideband mode external loss rate $\kappa_{+,\mathrm{ex}}$ &
$2\pi \cdot 108.5~\mathrm{MHz}$ \\ \hline
Optical red sideband mode total loss rate $\kappa_{+}=\kappa_{+,\mathrm{in}}+\kappa_{+,\mathrm{ex}}$ &
$2\pi \cdot 274.2~\mathrm{MHz}$ \\ \hline
Optical blue sideband mode frequency $\omega_{-}$ &
$2\pi \cdot 194.1012~\mathrm{THz}$ \\ \hline
Optical blue sideband mode intrinsic loss rate $\kappa_{-,\mathrm{in}}$ &
$2\pi \cdot 125.0~\mathrm{MHz}$ \\ \hline
Optical blue sideband mode external loss rate $\kappa_{-,\mathrm{ex}}$ &
$2\pi \cdot 135.0~\mathrm{MHz}$ \\ \hline
Optical blue sideband mode total loss rate $\kappa_{-}=\kappa_{-,\mathrm{in}}+\kappa_{-,\mathrm{ex}}$ &
$2\pi \cdot 260.0~\mathrm{MHz}$ \\ \hline
Microwave mode frequency $\omega_{\mathrm{m}}$ &
$2\pi \cdot 5.669~\mathrm{GHz}$ \\ \hline
Microwave mode intrinsic loss rate $\kappa_{\mathrm{m,in}}$ &
$2\pi \cdot 31.9~\mathrm{MHz}$ \\ \hline
Microwave mode external loss rate $\kappa_{\mathrm{m,ex}}$ &
$2\pi \cdot 31.8~\mathrm{MHz}$ \\ \hline
Microwave mode total loss rate $\kappa_{\mathrm{m}}=\kappa_{\mathrm{m,in}}+\kappa_{\mathrm{m,ex}}$ &
$2\pi \cdot 63.7~\mathrm{MHz}$ \\ \hline
Experimentally estimated $g_{\mathrm{eo}}$ &
$2\pi \cdot 272~\mathrm{Hz}$ \\ \hline

\multicolumn{1}{|c|}{M2M Parameters} & \\ \hline
Signal mode frequency $\omega_{\mathrm{s}}$ &
$2\pi \cdot 5.637~\mathrm{GHz}$ \\ \hline
Signal mode intrinsic loss rate $\kappa_{\mathrm{s,in}}$ &
$2\pi \cdot 75~\mathrm{kHz}$ \\ \hline
Signal mode external loss rate $\kappa_{\mathrm{s,ex}}$ &
$2\pi \cdot 102~\mathrm{kHz}$ \\ \hline
Signal mode total loss rate $\kappa_{\mathrm{s}}=\kappa_{\mathrm{s,in}}+\kappa_{\mathrm{s,ex}}$ &
$2\pi \cdot 177~\mathrm{kHz}$ \\ \hline
Idler mode frequency $\omega_{\mathrm{i}}$ &
$2\pi \cdot 7.339~\mathrm{GHz}$ \\ \hline
Idler mode intrinsic loss rate $\kappa_{\mathrm{i,in}}$ &
$2\pi \cdot 193~\mathrm{kHz}$ \\ \hline
Idler mode external loss rate $\kappa_{\mathrm{i,ex}}$ &
$2\pi \cdot 565~\mathrm{kHz}$ \\ \hline
Idler mode total loss rate $\kappa_{\mathrm{i}}=\kappa_{\mathrm{i,in}}+\kappa_{\mathrm{i,ex}}$ &
$2\pi \cdot 758~\mathrm{kHz}$ \\ \hline

\multicolumn{1}{|c|}{Qubit Parameters} & \\ \hline
Readout resonator frequency $\omega_{\mathrm{r}}$ &
$2\pi \cdot 7.339~\mathrm{GHz}$ \\ \hline
Readout resonator linewidth $\kappa_{\mathrm{r}}$ &
$2\pi \cdot 245~\mathrm{kHz}$ \\ \hline
Qubit frequency $\omega_{\mathrm{q}}$ &
$2\pi \cdot 3.402~\mathrm{GHz}$ \\ \hline
\end{tabular}
\end{center}

\newpage
\section{M2O device}
\begin{figure}[h]
    \centering
    \includegraphics[width=0.9\linewidth]{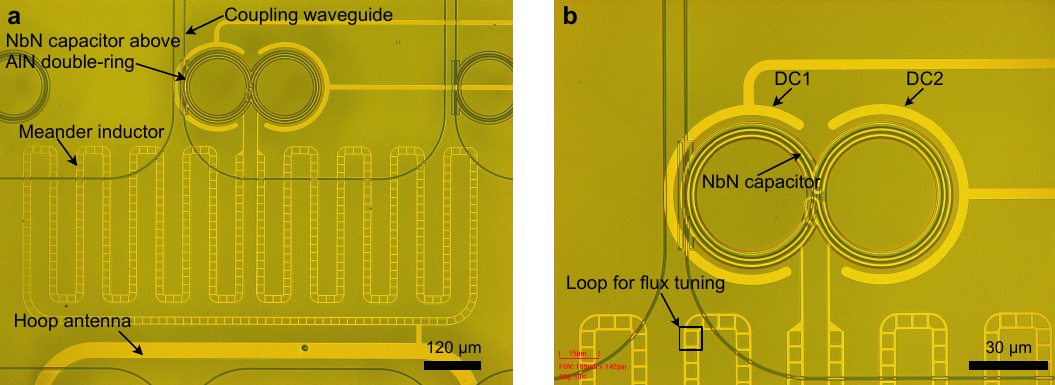}
    \caption{\textbf{Optical micrograph of the M2O device.} (a) The superconducting resonator is deposited above the AlN double-ring resonator. (b) Expanded view of the capacitor and dc electrodes of the superconducting resonator.}
    \label{fig:m2odevice}
\end{figure}

\newpage
\section{M2M device}
\begin{figure}[h]
    \centering
    \includegraphics[width=0.85\linewidth]{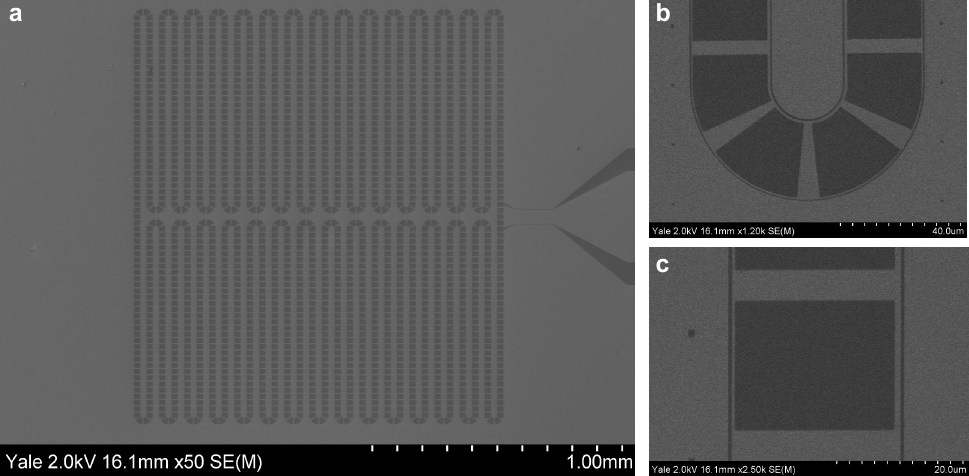}
    \caption{\textbf{SEM images of the M2M device.}
    (a) Overview SEM image of the M2M device, implemented as an enclosed meander resonator with integrated loop structures.
    The signal and pump tones are applied through a coupling pad that is capacitively coupled to the resonator.
    (b) Enlarged view of a device corner.
    (c) Enlarged view of the loop structure.}
    \label{fig:m2mdevice}
\end{figure}

The M2M device is shown in Fig.~\ref{fig:m2mdevice}. The device consists of an enclosed meander resonator with integrated loop structures. This geometry forms a distributed inductance and capacitance, supporting a dense set of multimode resonances. The loop structure, shown in Fig.~\ref{fig:m2mdevice}(c), incorporates two nanowires with intentionally asymmetric widths. When the device is biased with an external magnetic field, the resulting circulating supercurrent breaks the symmetry and enables flux-tunable frequency shifts, while also activating the three-wave-mixing interaction required for single-pump frequency conversion. For more detail of the design, simulation and device performance, please refer to Ref. \cite{wu2025broad}.


\newpage
\section{M2O noise calibration}
To calibrate the mode temperature of the M2O device’s microwave resonance, we insert a variable-temperature stage (VTS) upstream of the device. The VTS provides a calibrated thermal-noise reference, which allows us to extract the system gain and the added noise of the subsequent measurement chain. We then apply optical pulses to the transducer and infer the corresponding microwave-mode temperature following the procedure in Refs.~\cite{fu2021cavity, zhou2025kilometer}. The device-referred output noise spectrum can be written as
\begin{equation}
S_{\mathrm{dev}}(\omega)
= R(\omega)\bar{n}_{\mathrm{ex}}
+\bigl[1-R(\omega)\bigr]\bar{n}_{\mathrm{en}}, 
\end{equation}
where $R(\omega)$ is the measured power reflection spectrum of the superconducting resonator, $\bar{n}_{\mathrm{en}}$ and $\bar{n}_{\mathrm{ex}}$ are the thermal occupancies associated with the intrinsic (environment) bath and the external bath, respectively. The corresponding microwave-mode thermal occupancy is
\begin{equation}
\bar{n}_{\mathrm{mode}}
= \frac{\kappa_{m,\mathrm{in}}\,\bar{n}_{\mathrm{en}}+\kappa_{m,\mathrm{ex}}\,\bar{n}_{\mathrm{ex}}}
{\kappa_{m,\mathrm{in}}+\kappa_{m,\mathrm{ex}}}.
\label{equ:modePhoton}
\end{equation}

\begin{figure}[!htbp]
  \centering
  \includegraphics[width=0.5\linewidth]{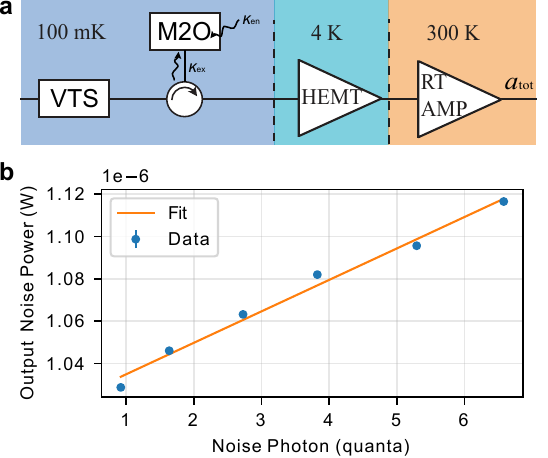}
  \caption{
    \textbf{Noise calibration of the M2O microwave mode using a variable-temperature stage (VTS).}
    (a) Schematic of the measurement chain: thermal noise generated by the VTS is injected into the M2O device at the MXC stage that is heated to \SI{100}{\milli\kelvin}, reflected from the microwave resonator, and amplified by a HEMT amplifier on the \SI{4}{\kelvin} stage followed by a room-temperature amplifier. 
    (b) Measured output noise power (points) versus the injected noise photon occupancy at the device input, together with a linear fit (line) used to extract the total system gain and the amplifier added noise referred to the VTS.}
  \label{fig:noisecalibration}
\end{figure}

As illustrated in Fig.~\ref{fig:noisecalibration}(a), noise from the VTS is sent to the M2O device, reflected by the microwave resonator, and routed to a high-electron-mobility transistor (HEMT) amplifier on the 4~K stage, followed by further amplification at room temperature. On resonance, the measured output noise power is related to the device-referred noise by
\begin{equation}
S_{\mathrm{out}} = G\bigl(S_{\mathrm{dev}} + n_{\mathrm{amp}}\bigr),
\end{equation}
where $G$ is the total system gain and $n_{\mathrm{amp}}$ is the total amplifier added noise, referred to the VTS input. We determine $G$ and $n_{\mathrm{amp}}$ by sweeping the VTS temperature (i.e., $\bar{n}_{\mathrm{ex}}$) and fitting the resulting $S_{\mathrm{out}}$ versus $\bar{n}_{\mathrm{ex}}$ data to a linear model, as shown in Fig.~\ref{fig:noisecalibration}(b). With this calibration, the light-induced mode temperature is obtained by comparing the output noise power with the laser on and off.

In this measurement, we apply an on-chip optical power of $0~\mathrm{dBm}$ using \SI{4}{\micro\second} pulses at a repetition rate of \SI{100}{\kilo\hertz}. Under these conditions, we extract a light-induced bath occupancy of $\bar{n}_{\mathrm{en}}=7.4\pm 0.4$ quanta, corresponding to a total microwave-mode occupancy of $\bar{n}_{\mathrm{mode}}=3.1 \pm 0.2$ quanta via Eq.~\eqref{equ:modePhoton}.

\newpage
\section{M2M conversion, amplification, and noise characterization}
\begin{figure}[h]
    \centering
    \includegraphics[width=\linewidth]{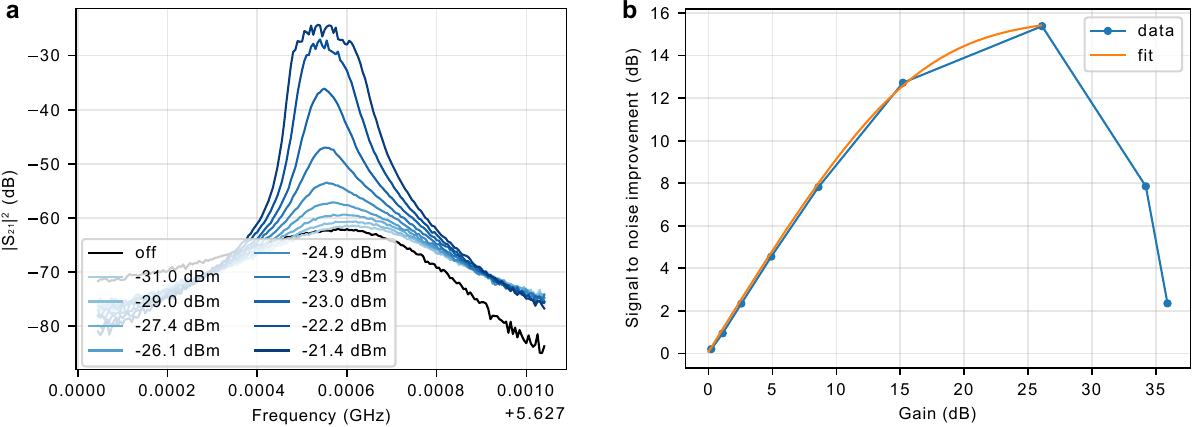}
    \caption{\textbf{Microwave frequency conversion with optional parametric amplification and resulting SNR benefit.}
    (a) Measured conversion response $|S_{21}|^{2}$ versus frequency. The black trace shows frequency conversion spectrum with the amplification pump off, while the blue traces show conversion plus amplification for the indicated pump powers (not calibrated on-chip).
    (b) Corresponding signal-to-noise ratio (SNR) improvement (relative to the pump-off case) plotted as a function of the achieved gain. The fitted line is to fit the added noise of the M2M device.}
    \label{fig:m2m_conversion_amp}
\end{figure}
In the optical qubit-readout demonstration, we operate the M2M device in both conversion and amplification modes. For frequency conversion, we pump at the difference frequency between the signal and idler modes. For amplification, we additionally apply a pump at twice the idler-mode frequency, and optimize the pump phase to achieve the maximum gain in the phase-sensitive amplification process. Fig.~\ref{fig:m2m_conversion_amp}(a) shows the measured spectra for pure conversion and for conversion with added amplification at several pump powers. Note that during these measurements the amplification pump frequency is swept together with the probe frequency; as a result, the observed amplification response does not follow a simple Lorentzian lineshape. As the amplification pump power is increased, the gain reaches up to $\sim 35$~dB. For gains above $\sim 30$~dB, however, the response begins to saturate, and spurious features emerge, producing a flat-top spectrum.

Figure~\ref{fig:m2m_conversion_amp}(b) shows the signal-to-noise ratio (SNR) enhancement as a function of the amplification gain. The SNR increases approximately linearly with gain up to $\sim 15$~dB, after which the improvement begins to saturate. The enhancement reaches a maximum of $\sim 15.5$~dB at a gain of about $25$~dB. At higher gain ($>25$~dB), the SNR enhancement decreases, indicating the onset of additional noise processes.

In this measurement, we could not place a variable temperature stage (VTS) ahead of the M2M device to directly calibrate the noise performance, because the VTS introduces a $30$~dB attenuation and would dissipate a prohibitive amount of heat at MXC \cite{xu2024radiatively}. To estimate the added noise of the overall chain, we model the system as an attenuator with power transmission $\eta$ between the M2M output and the input of the high-electron-mobility transistor (HEMT) amplifier. The HEMT is a Low Noise Factory LNF-LNC4\_16B with an average noise temperature of $3.1$~K. The SNR with and without M2M amplification can be written as
\begin{equation}
    \mathrm{SNR}_{\mathrm{wo}} = \frac{\eta P}{0.5 + n_\mathrm{HEMT}}, 
    \mathrm{SNR}_{\mathrm{w}} = \frac{\eta G P}{\eta G (0.5 + n_\mathrm{add}) + n_\mathrm{HEMT}}, 
\end{equation}
and the SNR enhancement can be written as a function of gain
\begin{equation}
\Delta \mathrm{SNR}
= \frac{G\left(0.5+n_{\mathrm{HEMT}}\right)}
{\eta\,G\left(0.5+n_{\mathrm{add}}\right)+n_{\mathrm{HEMT}}},
\end{equation}
where $P$ is the signal power, $G$ is the amplification gain, $n_{\mathrm{HEMT}}=10.9$ is the HEMT noise expressed in quanta at $5.66$~GHz, and $n_{\mathrm{add}}$ is the added noise referred to the M2M output. A two-parameter fit to Fig.~\ref{fig:m2m_conversion_amp}(b) yields $\eta\left(0.5+n_{\mathrm{add}}\right)=0.30$. Assuming an upper bound of $8$~dB insertion loss between the M2M device and the HEMT (i.e., $\eta>0.16$), we obtain an upper bound on the added noise of
$n_{\mathrm{add}}<1.3$ quanta.


\newpage
\section{Measurement circuits}
\begin{figure}[h]
    \centering
    \includegraphics[width=0.9\linewidth]{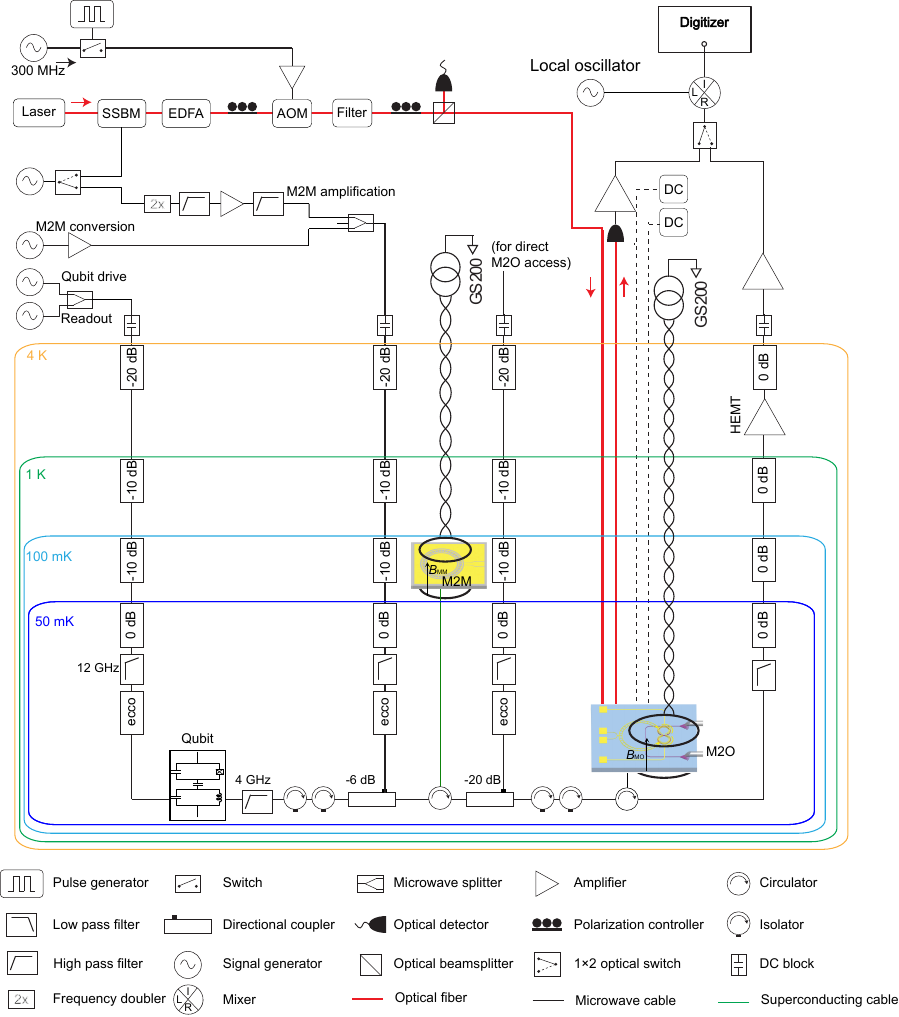}
    \caption{\label{fig:setup_appendix}
    Measurement setup for the cascaded M2M--M2O interface and qubit optical readout.}
\end{figure}


%% file: My_reference.bib
@article{zhou2025kilometer,
  title={A kilometer photonic link connecting superconducting circuits in two dilution refrigerators},
  author={Zhou, Yiyu and Wu, Yufeng and Li, Chunzhen and Shen, Mohan and Yang, Likai and Xie, Jiacheng and Tang, Hong X},
  journal={arXiv preprint arXiv:2508.02444},
  year={2025}
}

@article{xu2021bidirectional,
  title={Bidirectional interconversion of microwave and light with thin-film lithium niobate},
  author={Xu, Yuntao and Sayem, Ayed Al and Fan, Linran and Zou, Chang-Ling and Wang, Sihao and Cheng, Risheng and Fu, Wei and Yang, Likai and Xu, Mingrui and Tang, Hong X},
  journal={Nature communications},
  volume={12},
  number={1},
  pages={4453},
  year={2021},
  publisher={Nature Publishing Group UK London}
}

@article{andrews2014bidirectional,
  title={Bidirectional and efficient conversion between microwave and optical light},
  author={Andrews, Reed W and Peterson, Robert W and Purdy, Tom P and Cicak, Katarina and Simmonds, Raymond W and Regal, Cindy A and Lehnert, Konrad W},
  journal={Nature physics},
  volume={10},
  number={4},
  pages={321--326},
  year={2014},
  publisher={Nature Publishing Group UK London}
}

@article{fu2021cavity,
  title={Cavity electro-optic circuit for microwave-to-optical conversion in the quantum ground state},
  author={Fu, Wei and Xu, Mingrui and Liu, Xianwen and Zou, Chang-Ling and Zhong, Changchun and Han, Xu and Shen, Mohan and Xu, Yuntao and Cheng, Risheng and Wang, Sihao and others},
  journal={Physical Review A},
  volume={103},
  number={5},
  pages={053504},
  year={2021},
  publisher={APS}
}

@article{fan2018superconducting,
  title={Superconducting cavity electro-optics: a platform for coherent photon conversion between superconducting and photonic circuits},
  author={Fan, Linran and Zou, Chang-Ling and Cheng, Risheng and Guo, Xiang and Han, Xu and Gong, Zheng and Wang, Sihao and Tang, Hong X},
  journal={Science advances},
  volume={4},
  number={8},
  pages={eaar4994},
  year={2018},
  publisher={American Association for the Advancement of Science}
}

@article{xu2019frequency,
  title={Frequency-tunable high-Q superconducting resonators via wireless control of nonlinear kinetic inductance},
  author={Xu, Mingrui and Han, Xu and Fu, Wei and Zou, Chang-Ling and Tang, Hong X},
  journal={Applied Physics Letters},
  volume={114},
  number={19},
  year={2019},
  publisher={AIP Publishing}
}

@article{wu2025broad,
  title={Broad-spectrum coherent frequency conversion with kinetic inductance superconducting metastructures},
  author={Wu, Yufeng and Wang, Chaofan and Wang, Danqing and Xu, Mingrui and Zhou, Yiyu and Tang, Hong X},
  journal={Physical Review Applied},
  volume={24},
  number={2},
  pages={024015},
  year={2025},
  publisher={APS}
}

@article{wang2013quantum,
  title={Quantum state characterization of a fast tunable superconducting resonator},
  author={Wang, ZL and Zhong, YP and He, LJ and Wang, H and Martinis, John M and Cleland, AN and Xie, QW},
  journal={Applied Physics Letters},
  volume={102},
  number={16},
  year={2013},
  publisher={AIP Publishing}
}

@article{delaney2022superconducting,
  title={Superconducting-qubit readout via low-backaction electro-optic transduction},
  author={Delaney, RD and Urmey, MD and Mittal, S and Brubaker, BM and Kindem, JM and Burns, PS and Regal, CA and Lehnert, KW},
  journal={Nature},
  volume={606},
  number={7914},
  pages={489--493},
  year={2022},
  publisher={Nature Publishing Group UK London}
}

@article{arnold2025all,
  title={All-optical superconducting qubit readout},
  author={Arnold, Georg and Werner, Thomas and Sahu, Rishabh and Kapoor, Lucky N and Qiu, Liu and Fink, Johannes M},
  journal={Nature Physics},
  volume={21},
  number={3},
  pages={393--400},
  year={2025},
  publisher={Nature Publishing Group UK London}
}

@article{pan2025all,
  title={All-optical control and multiplexed readout of multiple superconducting qubits},
  author={Pan, Xiaoxuan and Ma, Chuanlong and Wang, Jia-Qi and Zhu, Zheng-Xu and Li, Linze and Chen, Jiajun and Yang, Yuan-Hao and Zhou, Yilong and Zou, Jia-Hua and Xu, Xin-Biao and others},
  journal={arXiv preprint arXiv:2512.21199},
  year={2025}
}

@article{van2025optical,
  title={Optical readout of a superconducting qubit using a piezo-optomechanical transducer},
  author={Van Thiel, TC and Weaver, MJ and Berto, F and Duivestein, P and Lemang, M and Schuurman, KL and {\v{Z}}emli{\v{c}}ka, M and Hijazi, F and Bernasconi, AC and Ferrer, C and others},
  journal={Nature Physics},
  volume={21},
  number={3},
  pages={401--405},
  year={2025},
  publisher={Nature Publishing Group UK London}
}

@article{lecocq2021control,
  title={Control and readout of a superconducting qubit using a photonic link},
  author={Lecocq, Florent and Quinlan, Franklyn and Cicak, Katarina and Aumentado, Jose and Diddams, SA and Teufel, JD},
  journal={Nature},
  volume={591},
  number={7851},
  pages={575--579},
  year={2021},
  publisher={Nature Publishing Group UK London}
}

@article{warner2025coherent,
  title={Coherent control of a superconducting qubit using light},
  author={Warner, Hana K and Holzgrafe, Jeffrey and Yankelevich, Beatriz and Barton, David and Poletto, Stefano and Xin, CJ and Sinclair, Neil and Zhu, Di and Sete, Eyob and Langley, Brandon and others},
  journal={Nature Physics},
  pages={1--8},
  year={2025},
  publisher={Nature Publishing Group UK London}
}

@article{han2020cavity,
  title={Cavity piezo-mechanics for superconducting-nanophotonic quantum interface},
  author={Han, Xu and Fu, Wei and Zhong, Changchun and Zou, Chang-Ling and Xu, Yuntao and Sayem, Ayed Al and Xu, Mingrui and Wang, Sihao and Cheng, Risheng and Jiang, Liang and others},
  journal={Nature communications},
  volume={11},
  number={1},
  pages={3237},
  year={2020},
  publisher={Nature Publishing Group UK London}
}

@article{rochman2023microwave,
  title={Microwave-to-optical transduction with erbium ions coupled to planar photonic and superconducting resonators},
  author={Rochman, Jake and Xie, Tian and Bartholomew, John G and Schwab, KC and Faraon, Andrei},
  journal={Nature Communications},
  volume={14},
  number={1},
  pages={1153},
  year={2023},
  publisher={Nature Publishing Group UK London}
}

@article{xie2025scalable,
  title={Scalable microwave-to-optical transducers at the single-photon level with spins},
  author={Xie, Tian and Fukumori, Rikuto and Li, Jiahui and Faraon, Andrei},
  journal={Nature Physics},
  pages={1--7},
  year={2025},
  publisher={Nature Publishing Group UK London}
}

@article{mckenna2020cryogenic,
  title={Cryogenic microwave-to-optical conversion using a triply resonant lithium-niobate-on-sapphire transducer},
  author={McKenna, Timothy P and Witmer, Jeremy D and Patel, Rishi N and Jiang, Wentao and Van Laer, Rapha{\"e}l and Arrangoiz-Arriola, Patricio and Wollack, E Alex and Herrmann, Jason F and Safavi-Naeini, Amir H},
  journal={Optica},
  volume={7},
  number={12},
  pages={1737--1745},
  year={2020},
  publisher={Optical Society of America}
}

@article{sahu2022quantum,
  title={Quantum-enabled operation of a microwave-optical interface},
  author={Sahu, Rishabh and Hease, William and Rueda, Alfredo and Arnold, Georg and Qiu, Liu and Fink, Johannes M},
  journal={Nature communications},
  volume={13},
  number={1},
  pages={1276},
  year={2022},
  publisher={Nature Publishing Group UK London}
}

@article{jiang2020efficient,
  title={Efficient bidirectional piezo-optomechanical transduction between microwave and optical frequency},
  author={Jiang, Wentao and Sarabalis, Christopher J and Dahmani, Yanni D and Patel, Rishi N and Mayor, Felix M and McKenna, Timothy P and Van Laer, Rapha{\"e}l and Safavi-Naeini, Amir H},
  journal={Nature communications},
  volume={11},
  number={1},
  pages={1166},
  year={2020},
  publisher={Nature Publishing Group UK London}
}

@article{mirhosseini2020superconducting,
  title={Superconducting qubit to optical photon transduction},
  author={Mirhosseini, Mohammad and Sipahigil, Alp and Kalaee, Mahmoud and Painter, Oskar},
  journal={Nature},
  volume={588},
  number={7839},
  pages={599--603},
  year={2020},
  publisher={Nature Publishing Group UK London}
}

@article{weaver2024integrated,
  title={An integrated microwave-to-optics interface for scalable quantum computing},
  author={Weaver, Matthew J and Duivestein, Pim and Bernasconi, Alexandra C and Scharmer, Selim and Lemang, Mathilde and Thiel, Thierry C van and Hijazi, Frederick and Hensen, Bas and Gr{\"o}blacher, Simon and Stockill, Robert},
  journal={Nature nanotechnology},
  volume={19},
  number={2},
  pages={166--172},
  year={2024},
  publisher={Nature Publishing Group UK London}
}

@article{higginbotham2018harnessing,
  title={Harnessing electro-optic correlations in an efficient mechanical converter},
  author={Higginbotham, Andrew P and Burns, PS and Urmey, MD and Peterson, RW and Kampel, NS and Brubaker, BM and Smith, G and Lehnert, KW and Regal, CA},
  journal={Nature Physics},
  volume={14},
  number={10},
  pages={1038--1042},
  year={2018},
  publisher={Nature Publishing Group UK London}
}

@article{arnold2020converting,
  title={Converting microwave and telecom photons with a silicon photonic nanomechanical interface},
  author={Arnold, G and Wulf, Matthias and Barzanjeh, Shabir and Redchenko, ES and Rueda, A and Hease, William J and Hassani, Farid and Fink, Johannes M},
  journal={Nature communications},
  volume={11},
  number={1},
  pages={4460},
  year={2020},
  publisher={Nature Publishing Group UK London}
}

@article{han2018coherent,
  title={Coherent microwave-to-optical conversion via six-wave mixing in Rydberg atoms},
  author={Han, Jingshan and Vogt, Thibault and Gross, Christian and Jaksch, Dieter and Kiffner, Martin and Li, Wenhui},
  journal={Physical review letters},
  volume={120},
  number={9},
  pages={093201},
  year={2018},
  publisher={APS}
}

@article{fernandez2019cavity,
  title={Cavity-enhanced Raman heterodyne spectroscopy in Er 3+: Y 2 SiO 5 for microwave to optical signal conversion},
  author={Fernandez-Gonzalvo, Xavier and Horvath, Sebastian P and Chen, Yu-Hui and Longdell, Jevon J},
  journal={Physical Review A},
  volume={100},
  number={3},
  pages={033807},
  year={2019},
  publisher={APS}
}

@article{yang2025multi,
  title={Multi-channel microwave-to-optics conversion utilizing a hybrid photonic-phononic waveguide},
  author={Yang, Yuan-Hao and Wang, Jia-Qi and Zhu, Zheng-Xu and Zeng, Yu and Li, Ming and Zhang, Yan-Lei and Lu, Juanjuan and Zhang, Qiang and Wang, Weiting and Dong, Chun-Hua and others},
  journal={arXiv preprint arXiv:2509.10052},
  year={2025}
}

@article{pintus2022ultralow,
  title={Ultralow voltage, high-speed, and energy-efficient cryogenic electro-optic modulator},
  author={Pintus, Paolo and Singh, Anshuman and Xie, Weiqiang and Ranzani, Leonardo and Gustafsson, Martin V and Tran, Minh A and Xiang, Chao and Peters, Jonathan and Bowers, John E and Soltani, Moe},
  journal={Optica},
  volume={9},
  number={10},
  pages={1176--1182},
  year={2022},
  publisher={Optica Publishing Group}
}

@article{holzgrafe2020cavity,
  title={Cavity electro-optics in thin-film lithium niobate for efficient microwave-to-optical transduction},
  author={Holzgrafe, Jeffrey and Sinclair, Neil and Zhu, Di and Shams-Ansari, Amirhassan and Colangelo, Marco and Hu, Yaowen and Zhang, Mian and Berggren, Karl K and Lon{\v{c}}ar, Marko},
  journal={Optica},
  volume={7},
  number={12},
  pages={1714--1720},
  year={2020},
  publisher={OSA}
}

@article{rueda2016efficient,
  title={Efficient microwave to optical photon conversion: an electro-optical realization},
  author={Rueda, Alfredo and Sedlmeir, Florian and Collodo, Michele C and Vogl, Ulrich and Stiller, Birgit and Schunk, Gerhard and Strekalov, Dmitry V and Marquardt, Christoph and Fink, Johannes M and Painter, Oskar and others},
  journal={Optica},
  volume={3},
  number={6},
  pages={597--604},
  year={2016},
  publisher={Optical Society of America}
}

@article{sahu2023entangling,
  title={Entangling microwaves with light},
  author={Sahu, Rishabh and Qiu, Liu and Hease, William and Arnold, Georg and Minoguchi, Yuri and Rabl, Peter and Fink, Johannes M},
  journal={Science},
  volume={380},
  number={6646},
  pages={718--721},
  year={2023},
  publisher={American Association for the Advancement of Science}
}

@article{pintus2022integrated,
  title={An integrated magneto-optic modulator for cryogenic applications},
  author={Pintus, Paolo and Ranzani, Leonardo and Pinna, Sergio and Huang, Duanni and Gustafsson, Martin V and Karinou, Fotini and Casula, Giovanni Andrea and Shoji, Yuya and Takamura, Yota and Mizumoto, Tetsuya and others},
  journal={Nature Electronics},
  volume={5},
  number={9},
  pages={604--610},
  year={2022},
  publisher={Nature Publishing Group UK London}
}

@article{zhu2020waveguide,
  title={Waveguide cavity optomagnonics for microwave-to-optics conversion},
  author={Zhu, Na and Zhang, Xufeng and Han, Xu and Zou, Chang-Ling and Zhong, Changchun and Wang, Chiao-Hsuan and Jiang, Liang and Tang, Hong X},
  journal={Optica},
  volume={7},
  number={10},
  pages={1291--1297},
  year={2020},
  publisher={Optical Society of America}
}

@article{shen2022coherent,
  title={Coherent coupling between phonons, magnons, and photons},
  author={Shen, Zhen and Xu, Guan-Ting and Zhang, Mai and Zhang, Yan-Lei and Wang, Yu and Chai, Cheng-Zhe and Zou, Chang-Ling and Guo, Guang-Can and Dong, Chun-Hua},
  journal={Physical Review Letters},
  volume={129},
  number={24},
  pages={243601},
  year={2022},
  publisher={APS}
}

@article{nicolas2023coherent,
  title={Coherent optical-microwave interface for manipulation of low-field electronic clock transitions in 171Yb3+: Y2SiO5},
  author={Nicolas, Louis and Businger, Moritz and Sanchez Mejia, T and Tiranov, Alexey and Chaneli{\`e}re, Thierry and Lafitte-Houssat, Elo{\"\i}se and Ferrier, Alban and Goldner, Philippe and Afzelius, Mikael},
  journal={npj Quantum Information},
  volume={9},
  number={1},
  pages={21},
  year={2023},
  publisher={Nature Publishing Group UK London}
}

@article{kimble2008quantum,
  title={The quantum internet},
  author={Kimble, H Jeff},
  journal={Nature},
  volume={453},
  number={7198},
  pages={1023--1030},
  year={2008},
  publisher={Nature Publishing Group}
}

@article{wehner2018quantum,
  title={Quantum internet: A vision for the road ahead},
  author={Wehner, Stephanie and Elkouss, David and Hanson, Ronald},
  journal={Science},
  volume={362},
  number={6412},
  pages={eaam9288},
  year={2018},
  publisher={American Association for the Advancement of Science}
}

@article{simon2017towards,
  title={Towards a global quantum network},
  author={Simon, Christoph},
  journal={Nature Photonics},
  volume={11},
  number={11},
  pages={678--680},
  year={2017},
  publisher={Nature Publishing Group UK London}
}

@article{devoret2013superconducting,
  title={Superconducting circuits for quantum information: an outlook},
  author={Devoret, Michel H and Schoelkopf, Robert J},
  journal={Science},
  volume={339},
  number={6124},
  pages={1169--1174},
  year={2013},
  publisher={American Association for the Advancement of Science}
}

@article{lauk2020perspectives,
  title={Perspectives on quantum transduction},
  author={Lauk, Nikolai and Sinclair, Neil and Barzanjeh, Shabir and Covey, Jacob P and Saffman, Mark and Spiropulu, Maria and Simon, Christoph},
  journal={Quantum Science and Technology},
  volume={5},
  number={2},
  pages={020501},
  year={2020},
  publisher={IOP Publishing}
}

@article{han2021microwave,
  title={Microwave-optical quantum frequency conversion},
  author={Han, Xu and Fu, Wei and Zou, Chang-Ling and Jiang, Liang and Tang, Hong X},
  journal={Optica},
  volume={8},
  number={8},
  pages={1050--1064},
  year={2021},
  publisher={Optical Society of America}
}

@article{vogt2019efficient,
  title={Efficient microwave-to-optical conversion using Rydberg atoms},
  author={Vogt, Thibault and Gross, Christian and Han, Jingshan and Pal, Sambit B and Lam, Mark and Kiffner, Martin and Li, Wenhui},
  journal={Physical Review A},
  volume={99},
  number={2},
  pages={023832},
  year={2019},
  publisher={APS}
}

@article{rigetti2012superconducting,
  title={Superconducting qubit in a waveguide cavity with a coherence time approaching 0.1 ms},
  author={Rigetti, Chad and Gambetta, Jay M and Poletto, Stefano and Plourde, Britton LT and Chow, Jerry M and C{\'o}rcoles, Antonio D and Smolin, John A and Merkel, Seth T and Rozen, Jim R and Keefe, George A and others},
  journal={Physical Review B—Condensed Matter and Materials Physics},
  volume={86},
  number={10},
  pages={100506},
  year={2012},
  publisher={APS}
}

@article{clerk2007using,
  title={Using a qubit to measure photon-number statistics of a driven thermal oscillator},
  author={Clerk, AA and Utami, D Wahyu},
  journal={Physical Review A—Atomic, Molecular, and Optical Physics},
  volume={75},
  number={4},
  pages={042302},
  year={2007},
  publisher={APS}
}

@article{li2025superfluid,
  title={Superfluid-cooled transmon qubits under optical excitation},
  author={Li, Chunzhen and Wu, Yufeng and Pace, Manuel CC and LaHaye, Matthew D and Senatore, Michael and Tang, Hong X},
  journal={PRX Quantum},
  volume={6},
  number={3},
  pages={030303},
  year={2025},
  publisher={APS}
}

@article{campbell2026transmon,
  title={Transmon Architecture for Emission and Detection of Single Microwave Photons},
  author={Campbell, Daniel L and McCoy, Stephen and Andrews, Melinda and Madden, Alexander and Horowitz, Viva R and Husremovi{\'c}, Bakir and Marash, Samuel and Nadeau, Christopher and Nguyen, Man and Senatore, Michael and others},
  journal={arXiv preprint arXiv:2601.11378},
  year={2026}
}

@article{miyamura2025generation,
  title={Generation of Frequency-Tunable Shaped Single Microwave Photons Using a Fixed-Frequency Superconducting Qubit},
  author={Miyamura, Takeaki and Sunada, Yoshiki and Wang, Zhiling and Ilves, Jesper and Matsuura, Kohei and Nakamura, Yasunobu},
  journal={PRX Quantum},
  volume={6},
  number={2},
  pages={020347},
  year={2025},
  publisher={APS}
}

@article{cheng2019superconducting,
  title={Superconducting nanowire single-photon detectors fabricated from atomic-layer-deposited NbN},
  author={Cheng, Risheng and Wang, Sihao and Tang, Hong X},
  journal={Applied Physics Letters},
  volume={115},
  number={24},
  year={2019},
  publisher={AIP Publishing}
}

@article{xu2024radiatively,
  title={Radiatively cooled quantum microwave amplifiers},
  author={Xu, Mingrui and Wu, Yufeng and Dai, Wei and Tang, Hong X},
  journal={Applied Physics Letters},
  volume={125},
  number={2},
  year={2024},
  publisher={AIP Publishing}
}
